%% file: point-source-paper.tex
\documentclass[
aps,%
12pt,%
final,%
notitlepage,%
oneside,%
onecolumn,%
nobibnotes,%
nofootinbib,%
superscriptaddress,%
noshowpacs,%
centertags]%
{revtex4}

\usepackage{hyperref}
\usepackage{graphicx}
\usepackage{amsmath}
\usepackage{longtable}

\begin{document}

\title{First constraints on point-like astrophysical sources using Baikal-GVD muon neutrino events}

\input{Dec2025_authors_paper_En_revtex4.tex}

\begin{abstract}
Baikal-GVD is a new-generation neutrino telescope under final construction stages in Lake Baikal, Russia.
With an instrumented volume already at 0.7 km$^3$, Baikal-GVD is currently the largest neutrino telescope in the Northern hemisphere.
A sub-degree angular resolution, made possible thanks to high purity of Baikal water, further enhances Baikal-GVD sensitivity to cosmic neutrino sources.
In this work, we employ track-like events collected from the partially completed detector between April 2019 and March 2024 to search for muon neutrino fluxes from 92 astrophysical objects of interest.
For this, a $\chi^2$-based track reconstruction method is used along with a cut-based analysis.
The analysis uses upward-going muons only, providing coverage for declinations between $-90^\circ$ and $+38^\circ$.
No significant excess has been found, so upper limits are reported.
The obtained limits are competitive with those set by ANTARES and KM3NeT.
We briefly comment on a possible low-significance indication of an excess from the direction of Westerlund~1.
This work sets a major milestone on the way to full-scale scientific exploitation of Baikal-GVD data.

\end{abstract}

\maketitle

\section{Introduction}
Neutrino astronomy has been in active development in recent decades.
While the IceCube detector \cite{IceCube} at the South Pole remains the largest neutrino detector ever built,
the need for deeper observations of the neutrino sky has led to the development of two more km$^3$-scale neutrino telescopes, Baikal-GVD \cite{BaikalGVD} and KM3NeT \cite{KM3NeT_LoI}.
Both detectors are located in the Northern hemisphere, complementing IceCube with high-sensitivity observations of the Southern neutrino sky using track-like events under manageable atmospheric-muon-background conditions.
Additionally, both KM3NeT and Baikal-GVD promise an improved angular resolution compared to IceCube, thanks to reduced light scattering in liquid water compared to ice, 
which also strengthen the point-like source sensitivity.

The cosmic high-energy neutrino flux, discovered by IceCube in 2013 \cite{IceCube_diffuse}, has been studied with increasingly high precision using various IceCube measurement channels  \cite{IceCube_diffuse_2020,IceCube_diffuse_HESE_2021,IceCube_diffuse_tracks_2022} and recently confirmed by Baikal-GVD \cite{Baikal_cascades,Baikal_diffuse_2025}. 
However it remains uncertain what source(s) produce the bulk of the diffuse flux.
The Milky Way galaxy as a whole has been recently detected at a $4.5 \sigma$ level by IceCube \cite{IceCube_Galactic_diffuse}.
The measurements indicate that the Galactic emission makes up O(10\%) of the all-sky high-energy neutrino flux.
Indications of a neutrino flux from the Galactic plane has also been reported based on Baikal-GVD cascade data \cite{Baikal_Galactic_median}.
However, individual Galactic sources of high energy neutrino remain unresolved.
So far, the only detections of localized neutrino sources with statistical significance reaching a $4 \sigma$ level are the blazar TXS~0506+056 \cite{IceCube_TXS0506} and the Seyfert galaxy NGC~1068 \cite{IceCube_NGC1068}.
But these two sources can only account for a very small fraction of the diffuse cosmic flux.
The KM3NeT collaboration has reported upper limits on neutrino sources using data from first 21 deployed strings in combination with ANTARES data \cite{ANTARES_KM3NeT_point_source_ICRC_2025}, somewhat improving the limits from the complete ANTARES dataset \cite{ANTARES_15yr_point_sources}.
More sensitive observations are needed to uncover the populations of neutrino sources responsible for the diffuse flux and study the Galactic emissions.

The Baikal Gigaton Volume Detector (Baikal-GVD), under construction in Lake Baikal, Russia,
currently consists of 4212 optical modules arranged on 117 vertical strings, occupying a water volume of about 0.7~km$^3$.
The detector is optimized for the measurement of neutrino fluxes in the energy range between roughly $1$~TeV and $10$~PeV.
Baikal-GVD has already published some first results on neutrino source searches using cascade-like events, including the detection of a high-energy cascade from the direction of TXS 0506+056 \cite{Baikal_TXS_cascade} and a search for directional associations with radio-bright blazars and some other sources \cite{Baikal_cascades_MNRAS_2023}.

Baikal-GVD previously reported an observation of the atmospheric neutrino flux using track-like events \cite{Baikal-GVD_track_paper}.
In this work we further employ the track-like event data from the partially completed Baikal-GVD detector to set first upper limits on point-like neutrino sources, focusing on a pre-selected list of objects.
The analysis uses upward-going tracks only, which limits the declination range to $\text{Dec}<+38^\circ$. It is primarily sensitive to muon neutrino and muon anti-neutrino fluxes.

This paper is organized as follows.
Section \ref{section2} provides a brief description of the Baikal-GVD detector design and construction status.
Section \ref{section3} describes the point-like search analysis and its results,
which are then discussed in Section \ref{section4}.
Finally, Section \ref{section5} concludes the paper.

\section{Baikal-GVD}
\label{section2}

The Baikal-GVD detector site is located in the Southern basin of Lake Baikal at 51$^\circ$\,46'\,N 104$^\circ$\,24'\,E, 3.6 km offshore and 1360~m deep.
The light absorption length in the deep lake water reaches 22~m, while light scattering length is of the order of 40~m \cite{Baikal_optical_water_properties, Baikal_optical_water_properties_2025}.
After accounting for the forward-peaked diagram of light scattering ($\text{cos}(\theta_{\text{scat}}) \approx 0.85$), the effective scattering length in the relevant wavelength range is typically between about 200~m and 480~m \cite{Baikal_optical_water_properties, Baikal_optical_water_properties_2025}.
Every winter the lake is covered with ice up to about 1~m thick, providing a solid platform convenient for massive detector deployment and maintenance operations.

\begin{figure}
  \centering
  \includegraphics[height=10.0cm]{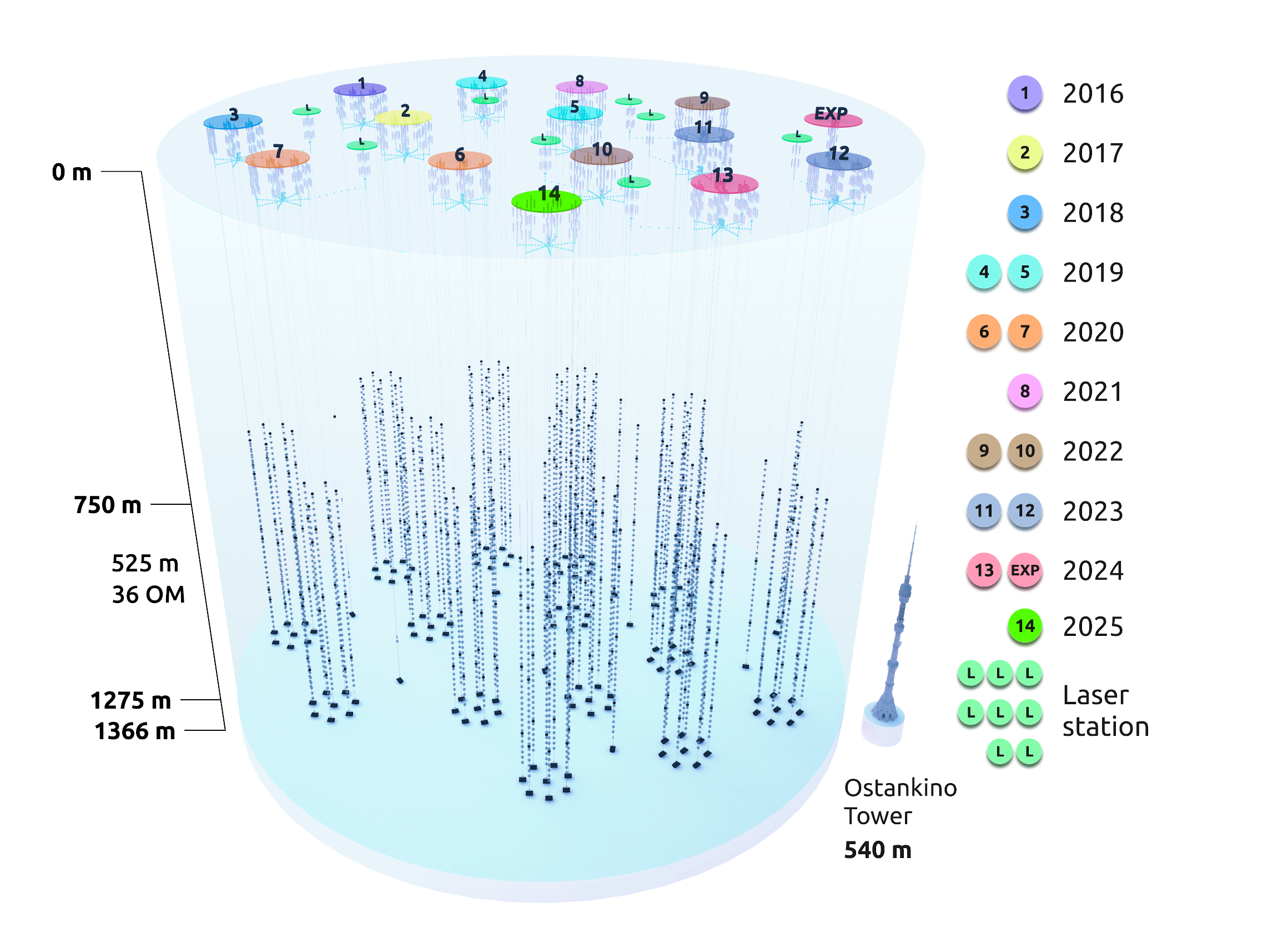}
  \caption{Schematic view of the Baikal-GVD detector. The legend shows the yearly progression of the detector construction.}
  \label{fig:baikal_gvd}
\end{figure}

The present Baikal-GVD layout is shown in Fig.~\ref{fig:baikal_gvd}. 
The optical module (OM) includes a 10-inch high-quantum-efficiency PMT (Hamamatsu R7081-100), a high voltage unit and front-end electronics, all within a pressure-resistant glass sphere.
The OMs are arranged on vertical strings.
Each string holds 36 OMs placed with a 15 m vertical spacing, at depths between 750~m and 1275~m below lake surface.
The strings also hold electronics modules responsible for OM signal digitization, slow control, power and data transmission, as well as hydrophones for acoustic positioning \cite{Baikal_positioning} and calibration LED beacons \cite{Baikal_calibration}.
The strings are grouped in clusters, 8 strings per cluster, with 60 m horizontal spacing between the strings within a cluster, and 250--300~m distance between cluster centers.
Additional strings equipped with laser beacons and, optionally, with OMs are installed in-between the GVD clusters.
The laser beacons are used for detector calibration and light propagation studies.
The detector has been taking data in partial configurations since 2016 when the first cluster was deployed.
Since then, one or two new clusters were added every year.
Currently, the instrument construction is 70\% complete, with 4212 optical modules installed on 117 vertical strings which are arranged in 14 clusters.
It is planned to reach a 20-cluster configuration (1 km$^3$) in 2028.
The data analysis presented in this paper uses data collected between April 2019 and March 2024, which includes clusters 1--11.

\section{Methods and results}
\label{section3}
We use a track-like event sample extracted from Baikal-GVD data collected between April 2019 and March 2024.
For this, we employ a $\chi^2$-based single-cluster track reconstruction method described in \cite{Safronov_et_al_ICRC_2025}, with minor improvements.
The analysis focuses on upward-going events ($\theta>90^\circ$) in order to suppress the large background of downward-going atmospheric muons.
The remaining background of mis-reconstructed atmospheric muons is further suppressed using a Boosted Decision Tree (BDT) event classifier, which combines various variables characterizing the reconstruction quality, event size and other related parameters in a single output variable (BDT score).
In this analysis we use a version of the BDT classifier 
optimized for muon energy above 10 TeV,
$BDT_{HE}$ \cite{Safronov_et_al_ICRC_2025}.
Based on Monte Carlo simulations, a cut at $BDT_{HE} > 0.25$ was found to optimize the point-like source sensitivity.
This event selection procedure provides an event sample with estimated contamination by atmospheric muons under 5\% \cite{Safronov_et_al_ICRC_2025}.
The obtained dataset consists of 988 neutrino candidate events, with energies between $\sim$~100~GeV and $\sim$~1~PeV. 
The dataset is dominated by atmospheric neutrino events, with median neutrino energy of about 1~TeV.
The typical angular resolution in this analysis is $\sim 0.5^\circ$, reaching $\approx 0.2^\circ$ for tracks with the longest visible lengths ($\approx 500$~m).

Using the track-like event sample, a guided search for point-like neutrino sources is carried out.
For this, we have compiled, based on subjective considerations, a list of astronomical objects of interest.
After requesting the sources to be in the field-of-view of our analysis ($\text{Dec} < +38.8^{\circ}$), the list consists of 92 objects (see Appendix).
This includes TXS 0506+056, NGC 1068, the published direction to the ultra-high-energy neutrino event KM3-230213A \cite{KM3-230213A_Nature}, the Galactic center, some objects with claimed associations to IceCube and ANTARES neutrino events and hotspots, and a number of VHE gamma-ray emitters, as well as some X-ray selected Seyfert galaxies and some other notable galactic and extragalactic objects.
This object list is intended to include all objects with previously claimed neutrino associations and other objects of top interest for neutrino astronomy, especially extragalactic ones, but it is not designed for completeness on any objective metric.
In particular, it misses many Galactic TeV gamma-ray sources and may miss some of the most recent claimed neutrino associations.
Note that objects with $\text{Dec} < -38.8^{\circ}$ are constantly within the field of view of this analysis.

The search was conducted in a search cone of a $2^{\circ}$ radius around each object, ignoring the sources' angular extensions and positional uncertainties.
It needs to be noted that the search cone radius was chosen so to optimize for best sensitivity, defined as median expected upper limit, to point-like sources with an $E^{-2}$ spectrum.
The cone radius is significantly larger than the typical angular resolution of the event sample; this is due to the relatively low background event count, which places the analysis in a low-statistics Poisson regime where tight background rejection cut are unnecessary.

\begin{figure}
 \centering
 \includegraphics[height=9.0cm,width=13.5cm]{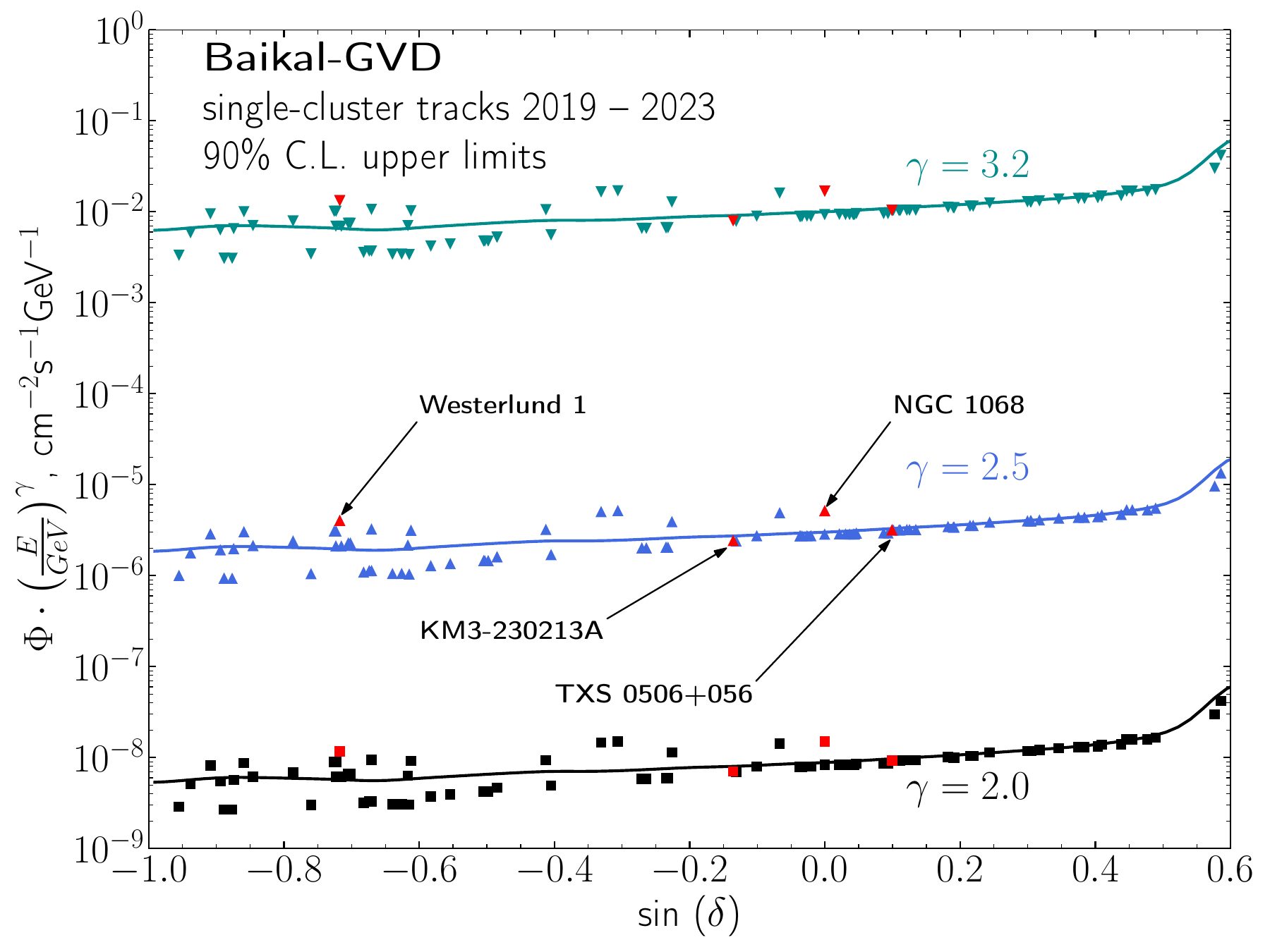}
 \caption{
  Neutrino point source sensitivity (median expected 90\% C.L. upper limit on per-flavour flux, solid color lines) and 90\% C.L. flux upper limits obtained for a catalogue of 92 source candidates using the Baikal-GVD track-like event sample collected between April 2019 to March 2024 (squares and triangles), for three assumed spectral indices $\gamma$.
  }
  \label{fig:track_analysis_results}
\end{figure}

The number of events found in the search cone was compared to a background estimate obtained from scrambling the data in right ascension.
No statistically significant excess has been found.
Upper limits for all objects in the list were calculated following the method of Feldman and Cousins \cite{Feldman_Cousins_1998}
under the assumption of a power law neutrino flux
with spectral indices $\gamma$ = 2; 2.5; 3.2 (see Fig.~\ref{fig:track_analysis_results} and Table~1).
The conversion of event counts to muon neutrino fluxes ($\Phi = dN/dE$) relies on Monte Carlo simulations of the Baikal-GVD detector \cite{BaikalGVD,Baikal-GVD_track_paper}.
The flux values refer to muon neutrino plus muon anti-neutrino flux
before attenuation in the Earth.
The attenuation of neutrino fluxes during their passage through Earth, including the effects of charged current and neutral current interactions, was computed using the NuFATE package \cite{NuFATE}.
The object with the largest number of neutrino candidate events falling in the $2^{\circ}$ degree
cone (three events) is the young massive star cluster and PeVatron candidate Westerlund~1 \cite{Westerlund_1_HESS_2022}.
The expected number of background events for this object is 0.98 events.
The corresponding pre-trial p-value for the background-only hypothesis is 0.0036, which corresponds to $2.7\,\sigma$ (pre-trial).

\section{Discussion}
\label{section4}
Our results are consistent with the background-only hypothesis.
The lack of significant detection of NGC~1068 is not surprising, considering that the total exposure in this analysis is an order of magnitude smaller than the IceCube accumulated exposure that led to the $4\,\sigma$ discovery of that source.
The same argument applies to TXS~0506+056, with the added uncertainty due to source variability.

Given the trial factor involved, the detection of three events from around Westerlund~1 can be and likely is an accidental coincidence.
Alternatively, it could be, at least in some part, due to neutrino emission from the Westerlund~1 region.
Indeed, young massive stellar clusters have been considered as potential sites of cosmic ray acceleration \cite{Parizot_2004} and Westerlund~1 is the most massive known young stellar cluster in our Galaxy \cite{Westerlund_1_mass}.
The 2$^\circ$-diameter TeV gamma-ray source HESS J1646$-$458 appears to be related to the cluster wind termination shock of Westerlund 1 \cite{Westerlund_1_HESS_2022,Bykov_2014}, with a potentially viable hadronic model interpretation.
Thus HESS J1646$-$458 could in principle contribute to the neutrino events observed by Baikal-GVD, although translating the Baikal-GVD neutrino count (three events, 
corresponding to $E^2 dN/dE = 5.5 \times 10^{-9} \, \text{GeV} \, \text{cm}^{-2} \, \text{s}^{-1}$), 
under a hadronic emission scenario, into gamma-ray flux would result in overshooting the HESS J1646$-$458 flux,
which is about $4 \times 10^{-9} \, \text{GeV} \, \text{cm}^{-2} \, \text{s}^{-1}$ at 10~TeV ($E^2 dN_\gamma/dE$).
Our 2$^\circ$-radius search region also includes two other known TeV gamma-ray sources, HESS J1640$-$465 and HESS J1641$-$463, which could potentially also be sites of hadronic processes \cite{Westerlund_1_Mares_2021}.
It is also possible that the region might host a more extended source which could escape detection by
H.E.S.S.
Indeed, Imaging Atmospheric Cherenkov Telescopes, such as H.E.S.S., are known to have difficulties observing sources with extensions comparable to or larger than their field of view (5$^\circ$ for H.E.S.S., see, e.g., \cite{HESS_Galactic_diffuse_2014}).
Note that this source is located too far South for the wide field-of-view observatories LHAASO and HAWC.
Finally, the region is located right on the Galactic plane, at $(l,b)=(339.55^\circ,-0.40^\circ)$, where the Galactic diffuse emission could provide a sizable contribution to the total flux.

The upper limits presented in this work do not include corrections for systematic uncertainties. 
Uncertainties may come from imperfect knowledge of the detector efficiency and angular resolution.
A recent analysis of the atmospheric neutrino spectrum using Baikal-GVD track data \cite{Safronov_et_al_ICRC_2025} suggests that the detector efficiency in Baikal-GVD track-like event simulations is under control within $\pm$35\%, with underprediction appearing more likely than overprediction. The upper limits reported here do not include any correction for the detector efficiency uncertainty.
Systematic uncertainties in the knowledge of the angular resolution are not expected to have a significant impact on the present analysis because of the relatively large size of the search cone ($2^\circ$) compared to the typical angular resolution ($<1.0^\circ$ at 68\% containment).

The results presented in this work make use of a limited Baikal-GVD dataset and simplistic event reconstruction and analysis techniques.
The exposure used corresponds to 30 cluster-years, or 1.5 yr of the full-detector (1 km$^3$) operation.
As one can see, the attained sensitivity is already similar to that reported for the ANTARES 15-yr dataset \cite{ANTARES_15yr_point_sources}.
The analysis sensitivity will further improve as more data are accumulated and the analysis techniques
are refined. 
A re-analysis of the same dataset using a likelihood-based method and incorporating multi-cluster events would already provide a sizable improvement in sensitivity.

\begin{figure}
 \centering
 \includegraphics[height=9cm]{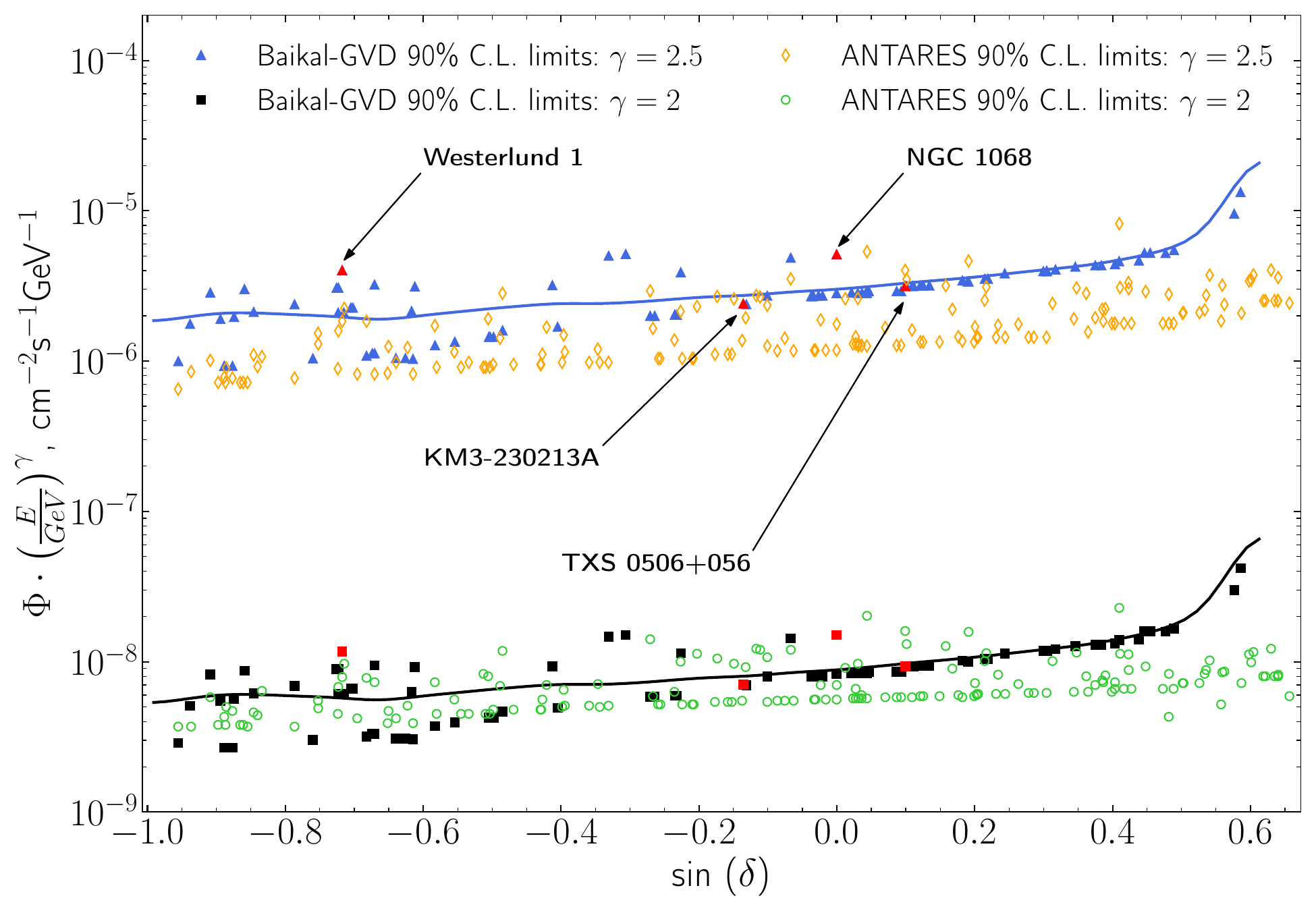}
 \caption{
  Comparison of the Baikal-GVD results (filled squares and triangles, 
  same as Fig.~\ref{fig:track_analysis_results}) with the ANTARES 15~yr limits from \cite{ANTARES_15yr_point_sources} (empty circles).
  }
  \label{fig:track_analysis_results_comparison}
\end{figure}

IceCube also possesses sensitivity to Southern hemisphere sources, which is strongly competitive with ANTARES, notably using high-energy cascade-like events \cite{IceCube_all_sky_point_sources_2025}.
Thus the flux limits reported in this work represent 
competitive
limits on neutrino flux from the listed point-like sources,
providing a sizable contribution to the total sensitivity of the worldwide neutrino telescope network.
The sensitivity of Baikal-GVD analyses will improve as the detector grows and collects more data and as the analysis techniques improve.
A likelihood-based point-source analysis of the Baikal-GVD data is currently in preparation.

\section{Conclusion}
\label{section5}
A first search for point-like neutrino sources with Baikal-GVD has been conducted.
The analysis uses single-cluster track-like events recorded by the partially completed Baikal-GVD detector within a 5~yr period starting April 2019.
The total livetime of the dataset is about 30 cluster-years, or, equivalently, 1.5 km$^3$-years.
After a strict event selection, the directions of the remaining events were compared to the directions of 92 astrophysical objects of interest using a simple cut-and-count approach, with a $2^\circ$ radius search cone around each object.
The analysis yielded no significant detection.
The object with the smallest pre-trial p-value in this analysis (0.0036) is the young star cluster Westerlund 1.
We report upper limits for each object on our list.
The limits are competitive with those set by ANTARES and KM3NeT.
The analysis sensitivity will improve as the detector grows, more data are collected, and the analysis techniques improve.

\section{Acknowledgements}
This work used data obtained with the Unique Scientific Installation "Baikal-GVD", operated within the Shared Research Center "Baikal Neutrino Observatory" of the Institute for Nuclear Research of the Russian Academy of Sciences.
This work is supported in the framework of the State project ``Science'' by the Ministry of Science and Higher Education of the Russian Federation under the contract 075-15-2024-541.

\clearpage
\appendix
\section{Tabulated results}

Table~1:
Upper limits on per-flavour neutrino flux from analysed astrophysical objects.  
Reported are the source name, declination $\delta$, right ascension $\alpha$, the number of events observed in the $2^\circ$-radius search cone, the pre-trial p-value of the background-only hypothesis and the obtained 90\% C.L. upper limits on $(E/GeV)^\gamma \, dN_{\nu_{\mu}}/dE$, in units of $10^{-8}$, $10^{-6}$ and $10^{-3} \, \text{cm}^{-2} \, \text{s}^{-1} \, \text{GeV}^{-1}$, under three different assumptions for the spectral index, $\gamma = 2$, 2.5 and 3.2, respectively.

\input{source_list_sorted.tex}

\clearpage

\end{document}

%% file: Dec2025_authors_paper_En_revtex4.tex

\author{V.A.~Allakhverdyan} 
\affiliation{Joint Institute for Nuclear Research, Dubna, 141980, Russia}
\author{A.D.~Avrorin}
\affiliation{Institute for Nuclear Research, Russian Academy of Sciences, Moscow, 117312, Russia}
\author{A.V.~Avrorin}
\affiliation{Institute for Nuclear Research, Russian Academy of Sciences, Moscow, 117312, Russia}
\author{V.M.~Aynutdinov}
\affiliation{Institute for Nuclear Research, Russian Academy of Sciences, Moscow, 117312, Russia}
\author{I.A.~Belolaptikov}
\affiliation{Joint Institute for Nuclear Research, Dubna, 141980, Russia}
\author{Z.Be\v{n}u\v{s}ov\'{a}}
\affiliation{Comenius University, Bratislava, 81499, Slovakia}
\affiliation{Czech Technical University in Prague, Prague, 11000, Czech Republic}
\author{E.A.~Bondarev}
\thanks{Corresponding authors}
\affiliation{Institute for Nuclear Research, Russian Academy of Sciences, Moscow, 117312, Russia}
\author{I.V.~Borina}
\affiliation{Joint Institute for Nuclear Research, Dubna, 141980, Russia}
\author{N.M.~Budnev}
\affiliation{Irkutsk State University, Irkutsk, 664003, Russia}
\author{V.A.~Chadymov}
\affiliation{Independent researchers}
\author{A.S.~Chepurnov}
\affiliation{Skobeltsyn Institute of Nuclear Physics MSU, Moscow, 119991, Russia}
\author{V.Y.~Dik}
\affiliation{Joint Institute for Nuclear Research, Dubna, 141980, Russia}
\affiliation{Institute of Nuclear Physics of the Ministry of Energy of the Republic of Kazakhstan, Almaty, 050032, Kazakhstan}
\author{A.N.~Dmitrieva}
\affiliation{National Research Nuclear University MEPHI, Moscow, 115409, Russia}
\author{G.V.~Domogatsky}
\thanks{Deceased}
\affiliation{Institute for Nuclear Research, Russian Academy of Sciences, Moscow, 117312, Russia}
\author{A.A.~Doroshenko}
\affiliation{Institute for Nuclear Research, Russian Academy of Sciences, Moscow, 117312, Russia}
\author{R.~Dvornick\'{y}}
\affiliation{Comenius University, Bratislava, 81499, Slovakia}
\affiliation{Czech Technical University in Prague, Prague, 11000, Czech Republic}
\author{A.N.~Dyachok}
\affiliation{Irkutsk State University, Irkutsk, 664003, Russia}
\author{Zh.-A.M.~Dzhilkibaev}
\affiliation{Institute for Nuclear Research, Russian Academy of Sciences, Moscow, 117312, Russia}
\author{E.~Eckerov\'{a}}
\affiliation{Comenius University, Bratislava, 81499, Slovakia}
\affiliation{Czech Technical University in Prague, Prague, 11000, Czech Republic}
\author{T.V.~Elzhov}
\affiliation{Joint Institute for Nuclear Research, Dubna, 141980, Russia}
\author{V.N.~Fomin}
\affiliation{Independent researchers}
\author{A.R.~Gafarov}
\affiliation{Irkutsk State University, Irkutsk, 664003, Russia}
\author{K.V.~Golubkov}
\affiliation{Institute for Nuclear Research, Russian Academy of Sciences, Moscow, 117312, Russia}
\author{A.R.~Gordeev}
\affiliation{Joint Institute for Nuclear Research, Dubna, 141980, Russia}
\author{T.I.~Gress}
\affiliation{Irkutsk State University, Irkutsk, 664003, Russia}
\author{K.G.~Kebkal}
\affiliation{LATENA, St. Petersburg, 199106, Russia}
\author{V.K.~Kebkal}
\affiliation{LATENA, St. Petersburg, 199106, Russia}
\author{I.V.~Kharuk}
\affiliation{Institute for Nuclear Research, Russian Academy of Sciences, Moscow, 117312, Russia}
\author{S.S.~Khokhlov}
\affiliation{National Research Nuclear University MEPHI, Moscow, 115409, Russia}
\author{E.V.~Khramov}
\affiliation{Joint Institute for Nuclear Research, Dubna, 141980, Russia}
\author{M.M.~Kolbin}
\affiliation{Joint Institute for Nuclear Research, Dubna, 141980, Russia}
\author{S.O.~Koligaev}
\affiliation{INFRAD, Dubna, 141980, Russia}
\author{K.V.~Konischev}
\affiliation{Joint Institute for Nuclear Research, Dubna, 141980, Russia}
\author{A.V.~Korobchenko}
\affiliation{Joint Institute for Nuclear Research, Dubna, 141980, Russia}
\author{A.P.~Koshechkin}
\affiliation{Institute for Nuclear Research, Russian Academy of Sciences, Moscow, 117312, Russia}
\author{V.A.~Kozhin}
\affiliation{Skobeltsyn Institute of Nuclear Physics MSU, Moscow, 119991, Russia}
\author{M.V.~Kruglov}
\affiliation{Joint Institute for Nuclear Research, Dubna, 141980, Russia}
\author{V.F.~Kulepov}
\affiliation{Nizhny Novgorod State Technical University, Nizhny Novgorod, 603950, Russia}
\author{A.A.~Kulikov}
\affiliation{Irkutsk State University, Irkutsk, 664003, Russia}
\author{Y.E.~Lemeshev}
\affiliation{Irkutsk State University, Irkutsk, 664003, Russia}
\author{M.V.~Lisitsin}
\affiliation{National Research Nuclear University MEPHI, Moscow, 115409, Russia}
\author{S.V.~Lovtsov}
\affiliation{Irkutsk State University, Irkutsk, 664003, Russia}
\author{R.R.~Mirgazov}
\affiliation{Irkutsk State University, Irkutsk, 664003, Russia}
\author{E.S.~Morgunov}
\affiliation{National Research Nuclear University MEPHI, Moscow, 115409, Russia}
\author{D.V.~Naumov}
\affiliation{Joint Institute for Nuclear Research, Dubna, 141980, Russia}
\author{A.S.~Nikolaev}
\affiliation{Skobeltsyn Institute of Nuclear Physics MSU, Moscow, 119991, Russia}
\author{I.A.~Perevalova}
\affiliation{Irkutsk State University, Irkutsk, 664003, Russia}
\author{A.A.~Petrukhin}
\affiliation{National Research Nuclear University MEPHI, Moscow, 115409, Russia}
\author{D.P.~Petukhov}
\affiliation{Institute for Nuclear Research, Russian Academy of Sciences, Moscow, 117312, Russia}
\author{E.N.~Pliskovsky}
\affiliation{Joint Institute for Nuclear Research, Dubna, 141980, Russia}
\author{M.I.~Rozanov}
\affiliation{St. Petersburg State Marine Technical University, St. Petersburg, 190008, Russia}
\author{E.V.~Ryabov}
\affiliation{Irkutsk State University, Irkutsk, 664003, Russia}
\author{G.B.~Safronov}
\affiliation{Institute for Nuclear Research, Russian Academy of Sciences, Moscow, 117312, Russia}
\author{B.A.~Shaybonov}
\affiliation{Joint Institute for Nuclear Research, Dubna, 141980, Russia}
\author{A.S.~Sheshukov} 
\affiliation{Joint Institute for Nuclear Research, Dubna, 141980, Russia}
\author{V.Y.~Shishkin}
\affiliation{Skobeltsyn Institute of Nuclear Physics MSU, Moscow, 119991, Russia}
\author{E.V.~Shirokov}
\affiliation{Skobeltsyn Institute of Nuclear Physics MSU, Moscow, 119991, Russia}
\author{F.~\v{S}imkovic}
\affiliation{Comenius University, Bratislava, 81499, Slovakia}
\affiliation{Czech Technical University in Prague, Prague, 11000, Czech Republic}
\author{A.E.~Sirenko}
\affiliation{Joint Institute for Nuclear Research, Dubna, 141980, Russia}
\author{A.V.~Skurikhin}
\affiliation{Skobeltsyn Institute of Nuclear Physics MSU, Moscow, 119991, Russia}
\author{A.G.~Solovjev}
\affiliation{Joint Institute for Nuclear Research, Dubna, 141980, Russia}
\author{M.N.~Sorokovikov}
\affiliation{Joint Institute for Nuclear Research, Dubna, 141980, Russia}
\author{I.~\v{S}tekl}
\affiliation{Czech Technical University in Prague, Prague, 11000, Czech Republic}
\author{A.P.~Stromakov}
\affiliation{Institute for Nuclear Research, Russian Academy of Sciences, Moscow, 117312, Russia}
\author{O.V.~Suvorova}
\affiliation{Institute for Nuclear Research, Russian Academy of Sciences, Moscow, 117312, Russia}
\author{V.A.~Tabolenko}
\affiliation{Irkutsk State University, Irkutsk, 664003, Russia}
\author{V.I.~Tretyak}
\affiliation{Joint Institute for Nuclear Research, Dubna, 141980, Russia}
\author{G.V.~Trubnikov}
\affiliation{Joint Institute for Nuclear Research, Dubna, 141980, Russia}
\author{B.B.~Ulzutuev}
\affiliation{Joint Institute for Nuclear Research, Dubna, 141980, Russia}
\author{Z.~Wang}
\affiliation{Joint Institute for Nuclear Research, Dubna, 141980, Russia}
\author{Y.V.~Yablokova}
\affiliation{Joint Institute for Nuclear Research, Dubna, 141980, Russia}
\author{D.N.~Zaborov}
\thanks{Corresponding authors}
\affiliation{Institute for Nuclear Research, Russian Academy of Sciences, Moscow, 117312, Russia}
\author{S.I.~Zavyalov}
\affiliation{Joint Institute for Nuclear Research, Dubna, 141980, Russia}
\author{D.Y.~Zvezdov}
\affiliation{Joint Institute for Nuclear Research, Dubna, 141980, Russia}

%% file: source_list_sorted.tex
\begin{longtable}{lccccccc}
\hline
Name	&	$\delta [^{\circ}]$	&	$\alpha [^{\circ}]$	&	$N_{obs}$	&	p-value	&	$\Phi^{90\% C.L.}_{\gamma = -2}$	&	$\Phi^{90\% C.L.}_{\gamma = -2.5}$	&	$\Phi^{90\% C.L.}_{\gamma = -3.2}$   \\ \hline
\endhead 
SMC	&	-72.80	&	13.16	&	0	&	1.0	&	0.29	&	1.0	&	3.36	\\ 
LMC	&	-69.76	&	80.89	&	1	&	2.6$\times 10^{-1}$	&	0.51	&	1.76	&	5.95	\\ 
Circinus Galaxy	&	-65.34	&	213.29	&	2	&	3.7$\times 10^{-2}$	&	0.82	&	2.85	&	9.61	\\ 
J1355-6326	&	-63.40	&	208.90	&	1	&	2.6$\times 10^{-1}$	&	0.55	&	1.9	&	6.41	\\ 
Cl Danks 1	&	-62.70	&	198.12	&	0	&	1.0	&	0.27	&	0.93	&	3.12	\\ 
Cl Danks 2	&	-62.68	&	198.22	&	0	&	1.0	&	0.27	&	0.93	&	3.12	\\ 
NGC 3603	&	-61.24	&	168.74	&	0	&	1.0	&	0.27	&	0.93	&	3.12	\\ 
Rabbit PWN	&	-60.98	&	214.52	&	1	&	2.6$\times 10^{-1}$	&	0.57	&	1.96	&	6.61	\\ 
ESO 138-G001	&	-59.23	&	252.83	&	2	&	3.7$\times 10^{-2}$	&	0.87	&	3.01	&	10.14	\\ 
Westerlund 2	&	-57.76	&	155.99	&	1	&	2.6$\times 10^{-1}$	&	0.61	&	2.12	&	7.14	\\ 
HESS J1614-518	&	-51.87	&	243.54	&	1	&	2.6$\times 10^{-1}$	&	0.69	&	2.38	&	8.01	\\ 
NGC 4945	&	-49.47	&	196.36	&	0	&	1.0	&	0.3	&	1.04	&	3.49	\\ 
HESS J1640-465	&	-46.53	&	250.18	&	2	&	3.7$\times 10^{-2}$	&	0.89	&	3.07	&	10.26	\\ 
Vela Jr	&	-46.33	&	133.00	&	1	&	2.6$\times 10^{-1}$	&	0.61	&	2.1	&	7.03	\\ 
HESS J1641-463	&	-46.30	&	250.26	&	2	&	3.7$\times 10^{-2}$	&	0.89	&	3.07	&	10.26	\\ 
{\bf Westerlund 1}	&	{\bf -45.85}	&	{\bf 251.76}	&	{\bf 3}	&	{\bf 3.6$\boldsymbol{\times 10^{-3}}$}	&	{\bf 1.17}	&	{\bf 4.01}	&	{\bf 13.42}	\\ 
Vela X	&	-45.66	&	128.88	&	1	&	2.6$\times 10^{-1}$	&	0.61	&	2.1	&	7.03	\\ 
PKS 1104-445	&	-44.82	&	166.79	&	1	&	2.6$\times 10^{-1}$	&	0.66	&	2.26	&	7.56	\\ 
PKS 2004-447	&	-44.58	&	301.98	&	1	&	2.6$\times 10^{-1}$	&	0.66	&	2.26	&	7.56	\\ 
Cen A	&	-43.02	&	201.37	&	0	&	1.0	&	0.32	&	1.08	&	3.62	\\ 
NGC 7582	&	-42.37	&	349.60	&	0	&	1.0	&	0.33	&	1.12	&	3.75	\\ 
HESS J1702-420A	&	-42.12	&	255.52	&	2	&	3.7$\times 10^{-2}$	&	0.95	&	3.22	&	10.74	\\ 
PKS B1424-418	&	-42.11	&	216.98	&	0	&	1.0	&	0.33	&	1.12	&	3.75	\\ 
RX J1713.7-3946	&	-39.76	&	258.39	&	0	&	1.0	&	0.31	&	1.04	&	3.46	\\ 
PKS 1954-388	&	-38.75	&	299.50	&	0	&	1.0	&	0.31	&	1.04	&	3.46	\\ 
NGC 424	&	-38.08	&	17.87	&	1	&	2.6$\times 10^{-1}$	&	0.63	&	2.15	&	7.13	\\ 
PKS 0426-380	&	-37.94	&	67.17	&	0	&	1.0	&	0.3	&	1.03	&	3.42	\\ 
NVSS J042025-374443	&	-37.75	&	65.10	&	2	&	3.7$\times 10^{-2}$	&	0.92	&	3.12	&	10.38	\\ 
PKS 1454-354	&	-35.65	&	224.36	&	0	&	1.0	&	0.37	&	1.27	&	4.24	\\ 
PKS 1313-333	&	-33.65	&	199.03	&	0	&	1.0	&	0.39	&	1.34	&	4.48	\\ 
IC 4329A	&	-30.31	&	207.33	&	0	&	1.0	&	0.42	&	1.45	&	4.83	\\ 
HESS J1745-303	&	-30.20	&	266.30	&	0	&	1.0	&	0.42	&	1.45	&	4.83	\\ 
M83	&	-29.87	&	204.25	&	0	&	1.0	&	0.42	&	1.45	&	4.83	\\ 
Galactic center	&	-29.01	&	266.42	&	0	&	1.0	&	0.47	&	1.59	&	5.32	\\ 
AP Librae	&	-24.37	&	229.42	&	1	&	2.6$\times 10^{-1}$	&	0.93	&	3.19	&	10.66	\\ 
5BZB J0630-2406	&	-23.89	&	97.56	&	0	&	1.0	&	0.49	&	1.69	&	5.64	\\ 
HESS J1809-193	&	-19.33	&	272.53	&	2	&	3.7$\times 10^{-2}$	&	1.47	&	5.01	&	16.73	\\ 
HESS J1813-178	&	-17.83	&	273.40	&	2	&	3.7$\times 10^{-2}$	&	1.51	&	5.14	&	17.19	\\ 
J0609-1542	&	-15.71	&	6.16	&	0	&	1.0	&	0.58	&	1.99	&	6.66	\\ 
PMN J1916-1519	&	-15.32	&	289.22	&	0	&	1.0	&	0.58	&	1.99	&	6.66	\\ 
TXS 0938-133	&	-13.60	&	145.26	&	0	&	1.0	&	0.59	&	2.03	&	6.77	\\ 
HESS J1825-137	&	-13.58	&	276.55	&	0	&	1.0	&	0.59	&	2.03	&	6.77	\\ 
HAWC J1825-134	&	-13.42	&	276.44	&	0	&	1.0	&	0.59	&	2.03	&	6.77	\\ 
NRAO 530	&	-13.08	&	263.26	&	1	&	2.6$\times 10^{-1}$	&	1.14	&	3.89	&	12.97	\\ 
KM3-230213A	&	-7.80	&	94.30	&	0	&	1.0	&	0.71	&	2.41	&	8.04	\\ 
TXS 2116-077	&	-7.54	&	319.72	&	0	&	1.0	&	0.7	&	2.38	&	7.95	\\ 
3C 279	&	-5.79	&	194.05	&	0	&	1.0	&	0.8	&	2.72	&	9.07	\\ 
PKS 1741-038	&	-3.83	&	266.00	&	1	&	2.6$\times 10^{-1}$	&	1.43	&	4.86	&	16.23	\\ 
3C 17	&	-2.13	&	9.59	&	0	&	1.0	&	0.79	&	2.7	&	9.01	\\ 
W40	&	-2.07	&	277.86	&	0	&	1.0	&	0.79	&	2.7	&	9.01	\\ 
W43	&	-1.94	&	281.88	&	0	&	1.0	&	0.79	&	2.7	&	9.01	\\ 
PKS 2332-017	&	-1.52	&	353.84	&	0	&	1.0	&	0.8	&	2.72	&	9.06	\\ 
PKS B0802-010	&	-1.19	&	121.30	&	0	&	1.0	&	0.8	&	2.72	&	9.06	\\ 
HESS J1849-000	&	-0.02	&	282.26	&	0	&	1.0	&	0.83	&	2.83	&	9.42	\\ 
NGC 1068	&	-0.01	&	40.67	&	1	&	2.6$\times 10^{-1}$	&	1.5	&	5.12	&	17.06	\\ 
W44	&	1.22	&	284.04	&	0	&	1.0	&	0.84	&	2.85	&	9.5	\\ 
PKS 0215+015	&	1.75	&	34.45	&	0	&	1.0	&	0.84	&	2.84	&	9.43	\\ 
HESS J1858+020	&	2.09	&	284.58	&	0	&	1.0	&	0.84	&	2.84	&	9.43	\\ 
NGC 6240	&	2.40	&	253.25	&	0	&	1.0	&	0.84	&	2.84	&	9.43	\\ 
3C 403	&	2.51	&	298.07	&	0	&	1.0	&	0.84	&	2.84	&	9.43	\\ 
HESS J1857+026	&	2.67	&	284.30	&	0	&	1.0	&	0.86	&	2.91	&	9.69	\\ 
SS 433	&	4.98	&	287.96	&	0	&	1.0	&	0.86	&	2.91	&	9.68	\\ 
3C 120	&	5.35	&	68.30	&	0	&	1.0	&	0.86	&	2.91	&	9.68	\\ 
TXS 0506+056	&	5.69	&	77.36	&	0	&	1.0	&	0.93	&	3.15	&	10.47	\\ 
HESS J0632+057	&	5.80	&	98.25	&	0	&	1.0	&	0.93	&	3.15	&	10.47	\\ 
GB6 J1040+0617	&	6.29	&	160.13	&	0	&	1.0	&	0.93	&	3.15	&	10.47	\\ 
MGRO J1908+06	&	6.39	&	287.05	&	0	&	1.0	&	0.93	&	3.15	&	10.47	\\ 
PKS 2145+067	&	6.96	&	327.02	&	0	&	1.0	&	0.94	&	3.18	&	10.55	\\ 
MAXI J1820+070	&	7.19	&	275.09	&	0	&	1.0	&	0.94	&	3.18	&	10.55	\\ 
PKS 2254+074	&	7.72	&	344.32	&	0	&	1.0	&	0.94	&	3.18	&	10.55	\\ 
TXS 1502+106	&	10.49	&	226.10	&	0	&	1.0	&	1.01	&	3.43	&	11.36	\\ 
GRS 1915+105	&	10.95	&	288.80	&	0	&	1.0	&	1.0	&	3.38	&	11.21	\\ 
PKS 0239+108	&	11.02	&	40.62	&	0	&	1.0	&	1.0	&	3.38	&	11.21	\\ 
M87	&	12.39	&	187.71	&	0	&	1.0	&	1.04	&	3.52	&	11.66	\\ 
NGC 4388	&	12.66	&	186.45	&	0	&	1.0	&	1.04	&	3.52	&	11.66	\\ 
W51	&	14.10	&	290.96	&	0	&	1.0	&	1.14	&	3.83	&	12.66	\\ 
PKS 2201+171	&	17.43	&	330.86	&	0	&	1.0	&	1.18	&	3.95	&	13.04	\\ 
PKS 0735+17	&	17.71	&	114.53	&	0	&	1.0	&	1.18	&	3.95	&	13.04	\\ 
PKS 1717+177	&	17.75	&	259.80	&	0	&	1.0	&	1.18	&	3.95	&	13.04	\\ 
J1826+1831	&	18.50	&	276.60	&	0	&	1.0	&	1.21	&	4.06	&	13.37	\\ 
RBS 0958	&	20.24	&	169.28	&	0	&	1.0	&	1.27	&	4.24	&	13.96	\\ 
Crab nebula	&	22.01	&	83.63	&	0	&	1.0	&	1.3	&	4.32	&	14.19	\\ 
IC 443	&	22.57	&	94.25	&	0	&	1.0	&	1.3	&	4.32	&	14.19	\\ 
PKS 1424+240	&	23.80	&	216.75	&	0	&	1.0	&	1.32	&	4.41	&	14.45	\\ 
MG3 J225517+2409	&	24.17	&	343.81	&	0	&	1.0	&	1.39	&	4.61	&	15.07	\\ 
NGC 4565	&	25.99	&	189.09	&	0	&	1.0	&	1.41	&	4.66	&	15.22	\\ 
3HWC J1951+266	&	26.50	&	298.15	&	0	&	1.0	&	1.59	&	5.24	&	17.05	\\ 
CGCG 164-019	&	27.03	&	221.40	&	0	&	1.0	&	1.59	&	5.24	&	17.05	\\ 
B2 2234+28A	&	28.48	&	339.09	&	0	&	1.0	&	1.59	&	5.23	&	17.01	\\ 
NGC 4278	&	29.28	&	185.03	&	0	&	1.0	&	1.66	&	5.46	&	17.71	\\ 
Cygnus X-1	&	35.20	&	299.59	&	0	&	1.0	&	2.99	&	9.55	&	30.42	\\ 
3HSP J095507.9+355101	&	35.85	&	148.78	&	0	&	1.0	&	4.19	&	13.29	&	42.13	\\ 
\hline
\end{longtable}